\documentclass[%
 reprint,
 amsmath,amssymb,
 aps,
]{revtex4-2}

\usepackage{graphicx}
\usepackage{dcolumn}
\usepackage{bm}
\usepackage{braket}
\usepackage{xcolor}
\usepackage{booktabs}
\usepackage[normalem]{ulem}

\usepackage{csquotes}

\begin{document}

\preprint{APS/123-QED}

\title{Continuous Accumulation of Cold Atoms in an Optical Cavity}




\author{Edward Gheorghita}
\thanks{Authors contributed equally to this work}
\author{Sebastian Wald}
\thanks{Authors contributed equally to this work}
\author{Andrea Pupi\'c}
\author{Onur Hosten}
\email{onur.hosten@ist.ac.at}
\affiliation{Institute of Science and Technology Austria (ISTA), 3400 Klosterneuburg, Austria}

\begin{abstract}
Continuously operating atom–light interfaces represent a key prerequisite for steady-state quantum sensors and efficient quantum processors. Here, we demonstrate continuous accumulation of sub-Doppler-cooled atoms in a shallow intracavity dipole trap, realizing this regime. The key ingredient is a light-shift manipulation that creates spatially varying cooling parameters, enabling efficient capture and accumulation of atoms within a cavity mode. Demonstrated with rubidium atoms, a continuous flux from a source cell is funneled through the magneto-optical trap into the cavity mode, where the atoms are cooled and maintained below $10~\mu\text{K}$ in steady state without time-sequenced operation. We characterize the resulting continuously maintained ensemble of millions of atoms and its collective coupling to the cavity field, establishing a route toward continuously operated cavity-QED systems and long-duration atomic and hybrid quantum sensors.
\end{abstract}

\maketitle

\emph{Introduction}---Continuous atom–light interfaces represent an important frontier for realizing quantum systems capable of uninterrupted sensing and efficient information processing. In such systems, the atomic and optical degrees of freedom can remain coherently coupled in steady state. Substantial progress has been made toward this goal in a range of platforms, including cavity-based quantum interfaces ~\cite{schafer2025,cline2025continuous,singh2022dual,chen2022,kwolek2022continuous,schioppo2017ultrastable,krauter2011entanglement}. However, achieving simultaneous atom loading, cooling, and confinement within an optical cavity remains a central experimental challenge due to conflicting requirements. The same light fields that provide trapping typically inhibit sub-Doppler cooling, preventing continuous accumulation of atoms. Consequently, such interfaces still rely on time-sequenced operation, alternating between loading, cooling, and interrogation phases.

Here, we demonstrate a method to overcome this challenge, continuously accumulating and cooling atoms directly within the mode of an optical cavity without time sequencing. This allows for maintaining a steady-state population of atoms continuously coupled to the cavity field, able to perform continuous interrogation. The key is a light-shift-manipulation technique that simultaneously allows capturing, cooling, and maintaining atoms in the same spatial region. Specifically, working with $^{87}$Rb atoms, we use a two-tone trap formed primarily by 1560-nm light with a small 1527-nm contribution. This configuration enables a simultaneous operation of a magneto-optical trap, a dipole trap, and optical molasses. Atoms are continuously transferred from the source to the magneto-optical trap (MOT) and then to the dipole trap. The required laser detunings for sub-Doppler cooling are naturally provided by the position-dependent energy shifts induced by the two-tone trap. In the process, several million atoms can accumulate in steady state, decoupled from the cooling light inside the trap, at temperatures that can reach 3~$\mu\text{K}$.

In a recent work, continuous atom–cavity coupling was achieved by cooling on a narrow optical transition and by employing spatially separated regions for cooling and coupling~\cite{cline2025continuous}. In contrast, our approach uses broad-line laser cooling to realize continuous loading and steady-state accumulation directly within the cavity mode, without relying on the naturally low temperatures available in narrow-line species.

The ability to sustain a steady-state ensemble of atoms within an optical cavity opens the way to novel and improved quantum protocols across a broad range of systems. These include continuous-wave active atomic clocks~\cite{reilly2025fully}, the possibility of steady-state operation for the emerging hybrid cold atom-optomechanical quantum systems~\cite{karg2020light}, and efficient reloading of atoms into a quantum processor for running
large-scale error-corrected quantum computations \cite{muniz2025repeated}.

In the following, we outline the contextual setting and the experimental setup, describe the tunable light shifts and the engineered continuous atom accumulation, and demonstrate the resulting continuous collective atom–cavity coupling.
\begin{figure*}[t]
  \centering
  \includegraphics[width=1\textwidth]{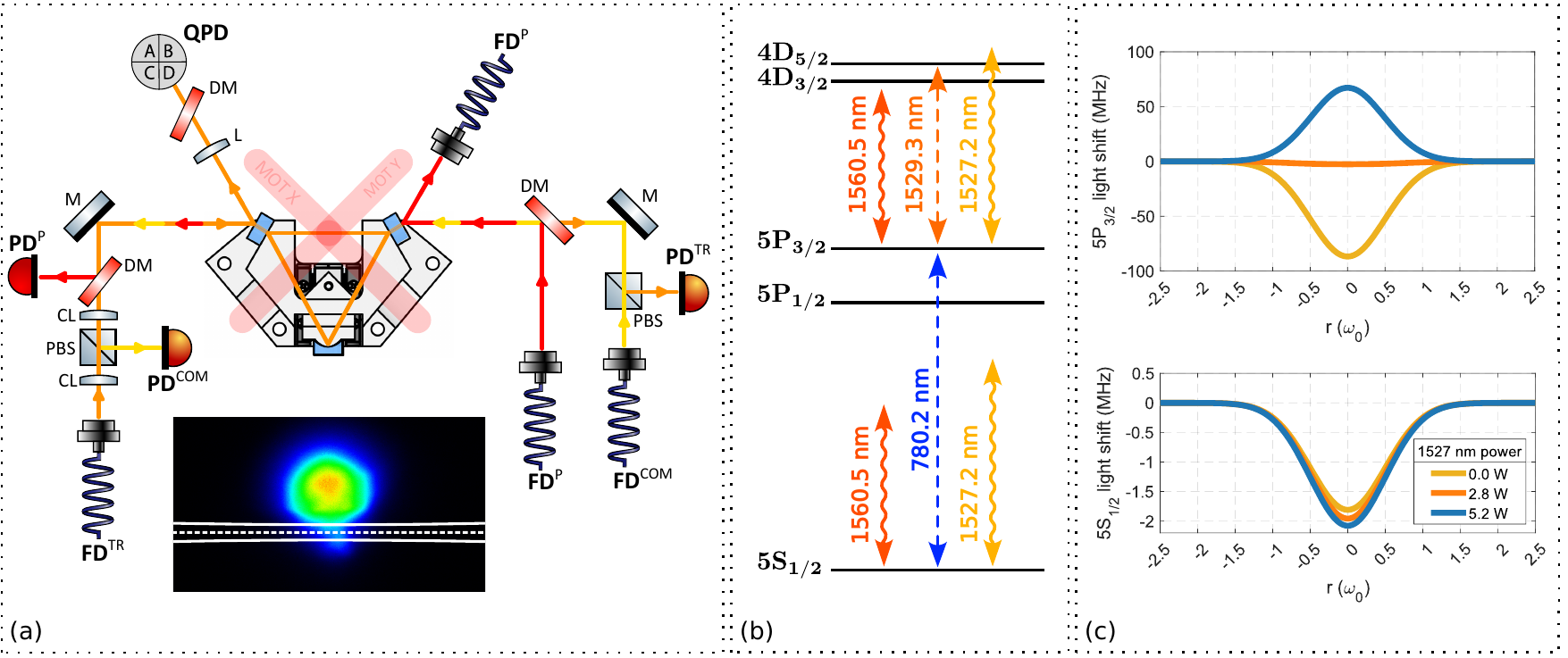}
  \caption{(a) Schematic of the traveling-wave cavity and the light tones coupled to the cavity formed by three mirrors (blue) attached to a monolithic spacer. The superscripts TR, COM and P refer to the trapping, compensation and cavity probe tones respectively. The 1560-nm trap tone is frequency locked to a high finesse cavity mode via the squash-locking method \cite{diorico2024laser} utilizing a quadrant photodiode (QPD) in reflection and the 780-nm probe tone is generated via frequency doubling. FD - fiber dock, PD - photodiode, M - mirror, DM - dichoric mirror, PBS - polarizing beam splitter, CL - cylindrical lens, L - spherical lens. Inset: Illustration of the geometry, where atoms are `\emph{sipped}' into the cavity mode (solid/dashed lines) from the edge of the 500-$\mu$m $1/e^2$-radius MOT, imaged here through its fluorescence in absence of the compensation tone. (b) Energy diagram for $^{87}\text{Rb}$, illustrating the relevant transition wavelengths (dashed) in connection with the applied tones. (c) Illustration of the total light-induced energy level shifts in the ground and excited states for three different 1527-nm intravacity powers for 157-$\mu\text{m}$ beam waist and 36-W intracavity 1560-nm power.}
  \label{fig:setup}
\end{figure*}

\emph{Contextual setting}---The setup utilized for this work was designed to explore entanglement-enhanced sensing~\cite{pezze2018quantum,szigeti2021improving}, with an emphasis on generating spin squeezed states \cite{kitagawa1993squeezed,schleier2010squeezing,cox2016,hosten2016measurement,Hosten2016Science}, enhancing \cite{malia2022distributed,greve2022entanglement,cassens2024entanglement} atom interferometric inertial sensors~\cite{cronin2009optics, ockeloen2013quantum,peters1999measurement,kasevich1991atomic}, and improving the performance of cavity-based atomic clocks~\cite{aeppli2024clock,pedrozo2020entanglement}. At the core of the setup is an in-vacuum traveling wave optical cavity (Fig.~\ref{fig:setup}a). The mirrors for the 3.05-GHz free-spectral-range cavity are coated for operation near both 780 nm and 1560 nm. Because the intracavity light undergoes reflections at non-normal incidence, the two polarizations can acquire distinct characteristics. The coatings are optimized to form a high finesse cavity for the out-of-plane polarization modes ($\mathcal{F}^\text{high}_{780}=39\times 10^3$ and $\mathcal{F}^\text{high}_{1560}=103\times 10^3$), while maintaining as low a finesse as possible for the in-plane polarization modes ($\mathcal{F}^\text{low}_{780}=2.2\times 10^3$ and $\mathcal{F}^\text{low}_{1560}=30\times 10^3$). The 780-nm modes interact with the atoms near-resonantly, while the 1560-nm modes form a far off-resonant dipole trap for the atoms. A 3D-MOT is nearly-overlaid with the cavity mode waist to load atoms from a commercial 2D-MOT-based atom source (see App.~\ref{app: setup}).

The 780/1560-nm dual-wavelength configuration offers both technical and fundamental advantages. It benefits from the low-noise infrastructure available in the telecom band around 1560 nm, allows streamlined frequency locking, in which a cavity-locked 1560-nm trap beam renders its frequency-doubled 780-nm tone resonant with the cavity, and enables uniform atom–cavity coupling in a standing-wave geometry \cite{hosten2016measurement}. Despite these desirable features, a 1560-nm tone rather strongly couples the 1529-nm $5\text{P}_{3/2}\leftrightarrow4\text{D}$ transitions of rubidium (Fig.~\ref{fig:setup}b). This coupling induces spatially varying light shifts on the $5\text{P}_{3/2}$ excited states with a magnitude nearly 50 times those for the $5\text{S}_{1/2}$ ground states (Fig.~\ref{fig:setup}c). This mismatch in light shifts poses challenges for operating a MOT, transferring atoms from the MOT to the dipole trap, and further executing well established cooling techniques. 

Overcoming these challenges has been a subject of investigation. The most straightforward operation for further cooling inside a 1560-nm dipole trap was to intermittently switch off the dipole trap ~\cite{engelsen2016quantum}. Furthermore, to improve loading efficiency into these traps, a narrow-resonance dark-state cooling method was investigated, nearing an order of magnitude improvement in the number of trapped atoms---reaching 1-2\% MOT-to-dipole trap transfer efficiencies and 200 $\mu$K in-trap temperatures \cite{naik2020loading,kuyumjyan2017condensation}. Building on the light-shift engineering ideas of Ref.~\cite{griffin2006spatially} for loading cold atoms into deep dipole traps, adding an auxiliary 1529-nm tone to a free-space 1560-nm dipole trap was shown to increase trapped atom numbers---reaching 7\% MOT-to-dipole-trap transfer efficiency and 50 $\mu$K in-trap temperatures \cite{Coop17,palacios2018multi}. Related 1529-nm light-shift control has also been employed to improve the uniformity of nondestructive atomic-state probing~\cite{vanderbruggen2013feedback,kohlhaas2015phase} and to spectrally hide atoms during state manipulations in optical tweezer arrays~\cite{bluvstein2025fault}.

As we demonstrate here, light shift compensation can not only improve transfer efficiencies and cooling beyond what has been shown, but can enable a conceptually distinct steady-state accumulation of atoms in a time-independent trap without interfering with atomic state control.

In our setup, a weak 1527-nm tone is used for light shift compensation. This tone is frequency locked to a low-finesse cavity mode with the side-of-fringe technique \cite{metcalf1999laser} to directly stabilize its intracavity optical power. The cavity modes at 1527 nm share essentially the same characteristics as those at 1560 nm. The power of the trap tone is additionally stabilized by monitoring the cavity transmission to achieve a low-noise trapping potential. Due to the power enhancement provided by the cavity, we only work with milliwatts of light for both tones, which turn into several Watts inside the cavity. We use two complementary methods to extract information from the trapped atoms: fluorescence measurements and cavity shift measurements. The fluorescence measurements take place either after releasing the atoms from the dipole trap or within it using light shift compensation. This is assisted by 780-nm MOT beams. The cavity shift measurements are nondestructive and provide real-time information  on atoms in the cavity mode. The 780-nm high-finesse cavity mode dispersively couples the $F=1$ and $F=2$ hyperfine states of the $5\text{S}_{1/2}$ ground state to the $5\text{P}_{3/2}$ excited states with opposite detunings ($\pm 3.4~\text{GHz}$), causing opposite cavity frequency shifts \cite{schleier2010squeezing,hammerer2004light} (see App.~\ref{app: parameters}). The total cavity shift, detected by scanning a probe beam to find the reflection dip frequency, measures the population difference between the two hyperfine ground states. If the total population is known to be only in one of the two states, it indicates the total atom number.

\emph{Control of light shifts}---The light shift of the $5\text{P}_{3/2}$ level is selectively controlled by varying the power of the auxiliary 1527-nm traveling-wave tone. In terms of the frequency-dependent scalar polarizability $\alpha_i$, the shift for level~\emph{i} can be expressed as $\Delta E_i = -\tfrac{1}{2}\alpha_i\mathcal{E}_\text{rms}^2$, where \mbox{$\mathcal{E}_\text{rms}$} is the position-dependent root-mean-square electric field of a single light tone \cite{arora2007magic,bernon2011trapping}. Due to the opposite detunings of the 1560-nm and 1527-nm tones for this transition, the corresponding polarizabilities carry opposite signs (Table~\ref{tab:table1}). Achieving a net-zero light shift on the $5\text{P}_{3/2}$ levels requires satisfying an intensity ratio of $I_{1527} = \tfrac{1}{12.1} I_{1560}$. In Fig.~\ref{fig:setup}(c), we show the calculated light shifts for three intensity ratios between the two tones, taking into account the spatial profile of the dipole trap and using the intensity relation $I = \epsilon_0 c \mathcal{E}_\text{rms}^2$.
\begin{figure}[t]
  \centering
  \includegraphics[width=0.95\columnwidth]{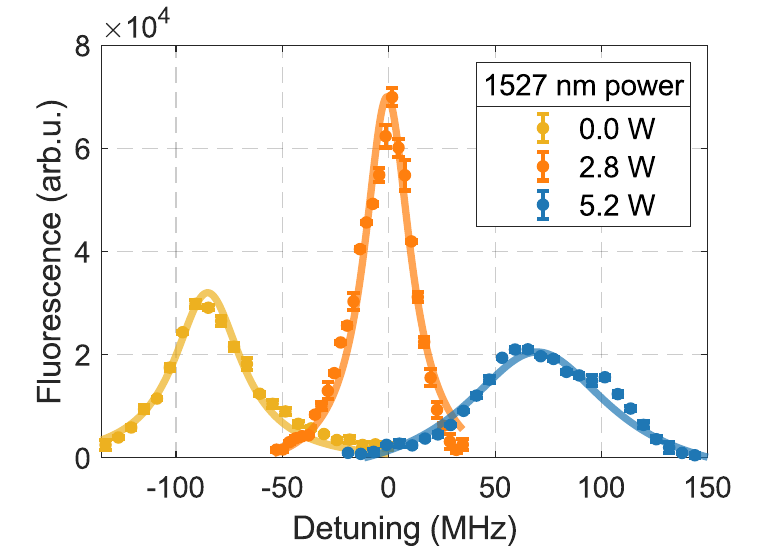}
  \caption{Characterization of light shift distributions for the \mbox{$F = 2 \rightarrow F' = 3$} transition for three different 1527 nm intracavity powers: no compensation (yellow), full compensation (orange) and overcompensation (blue). The 1560 nm intracavity power is fixed at 36~W. }
  \label{fig:energy_shifts}
\end{figure}

Atoms confined in the dipole trap experience light shifts in both the ground ($5\text{S}_{1/2}$) and the first excited ($5\text{P}_{3/2}$) states. A probe beam coupling these levels, with a scanned detuning, enables measurement of the differential light shift \mbox{$\delta=\Delta E_{5\text{P}_{3/2}}-\Delta E_{5\text{S}_{1/2}}$} via the resulting fluorescence signal \cite{brantut2008light,shih2013nondestructive}. In our implementation, the 780-nm MOT cooling beams ($5\text{S}_{1/2}, F=2 \leftrightarrow  5\text{P}_{3/2}, F'=3$) act as the probe, allowing characterization of the degree of light-shift compensation achieved for different compensation powers as illustrated in Figure~\ref{fig:energy_shifts}. During these 300-$\mu\text{s}$ measurements, both the trapping and compensation beams remain on, and the repump detuning is independently optimized for each compensation power to maximize fluorescence. The probe operates at around one saturation intensity, yielding a 10 MHz full-width for the fluorescence spectrum in absence of trap and compensation beams---smaller than the widths of the distributions in Fig.~\ref{fig:energy_shifts}. The mean shifts obtained for the three compensation powers demonstrate that we can shift the excited state levels to either direction in frequency in comparison to their natural values. The chosen intermediate compensation power cancels the mean differential shift caused by the trap. For a discussion on the widths of the light shift distributions and potential role of tensor shifts see App.~\ref{app: shifts}.
\begin{table}[b]
\caption{\label{tab:table1}%
Calculated scalar polarizabilities $\alpha_i$ in units $[\text{J}
][\text{m}^2]/[\text{W}]$. For $5\text{S}_{1/2}$, we include the contribution from the $5\text{S}_{1/2} \rightarrow (5\text{P}_{3/2},5\text{P}_{1/2})$ transitions. For $5\text{P}_{3/2}$, we include the contribution from the $5\text{P}_{3/2} \rightarrow (6\text{S}_{1/2},4\text{D}_{1/2},4\text{D}_{5/2})$ transitions \cite{bernon2011trapping}.}
\begin{ruledtabular}
\begin{tabular}{ccc}
$\lambda$ [nm] & $\alpha_{5\text{S}_{1/2}}$& $\alpha_{5\text{P}_{3/2}}$ \\
\midrule
1560 & $+6.804\times 10^{-39}$ & $+3.259\times 10^{-37}$ \\
1527 & $+6.906\times 10^{-39}$ & $-3.928\times 10^{-36}$ \\
\end{tabular}
\end{ruledtabular}
\end{table}

\emph{Continuous accumulation}---For a given dipole trap depth, beyond a threshold compensation power, we observe that atoms naturally start accumulating inside the dipole trap concurrently with MOT loading. The efficiency of this transfer does \emph{not} seem to depend critically on a good spatial overlap between the MOT and the cavity mode, as loading through the edge of the MOT (Fig.~\ref{fig:setup}a:inset) and through its center yields similar results.

\begin{figure}[t]
  \centering
  \includegraphics[width=0.95\columnwidth]{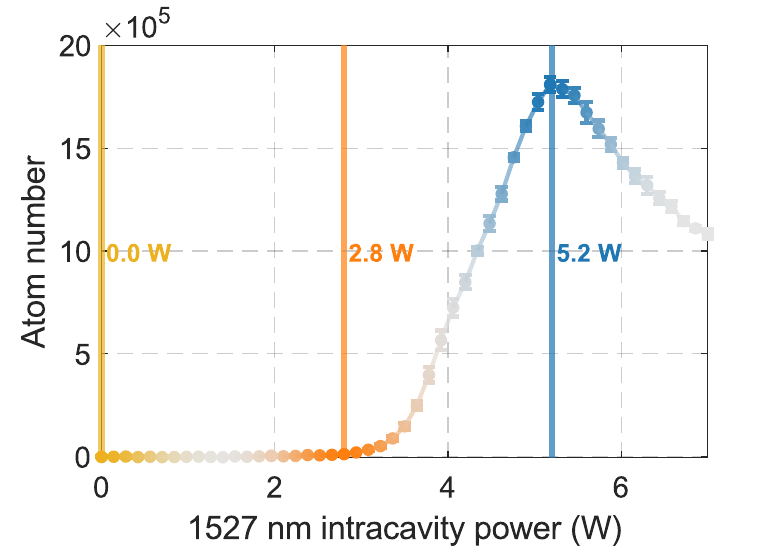}
  \caption{Number of accumulated atoms in the dipole trap after a 550-ms accumulation time as a function of compensation tone power. Three specific compensation levels associated with regions with qualitatively different behaviors are highlighted.}
  \label{fig:cont_load_atom_number}
\end{figure}
  
The number of atoms loaded into the 87-$\mu$K-deep (36~W) dipole trap we work with is demonstrated in Figure~\ref{fig:cont_load_atom_number} as a function of the 1527-nm intracavity power for a chosen accumulation duration of 550 ms. For reference, we highlight the three specific powers corresponding to the light shifts indicated and studied in Figs.~\ref{fig:setup}(c)~and~\ref{fig:energy_shifts}. First, in the absence of a compensation tone, the light shift from the 1560 nm locally inhibits the operation of the MOT. In this configuration, a MOT overlapped with the cavity mode can even fail to load atoms at sufficiently high trap powers. Next, at around 2.8~W in the compensation tone, while the MOT loads virtually unimpeded, no atoms are loaded into the dipole trap. Although this power corresponds to the intensity ratio at which the light shift is optimally compensated, the atomic temperatures in the MOT are too high to retain them in the utilized trap depth. Finally, for larger compensation powers (effectively red-detuning the cooling light for the atoms), a sharp rise in the number of atoms in the dipole trap is observed, reaching values as large as $1.8\times 10^6$ at the 550-ms accumulation time. This is 63\% of the full $2.9\times 10^6$ atoms accumulated in the dipole trap in steady state for this configuration.

These experiments begin by switching on the MOT coils as well as the cooling/repumper beams ($5~\text{mW}/\text{cm}^{2}$ per cooling beam, $180~\mu\text{W}/\text{cm}^{2}$ repumper), and setting a 1527 nm intracavity power. After 550 ms, the cooling/repumper beams and coils are turned off. Atoms are kept in the trap for another 100 ms before they are released from the trap for fluorescence imaging with a 300-$\mu\text{s}$ resonant pulse.

\begin{figure}[t]
  \centering
  \includegraphics[width=1\columnwidth]{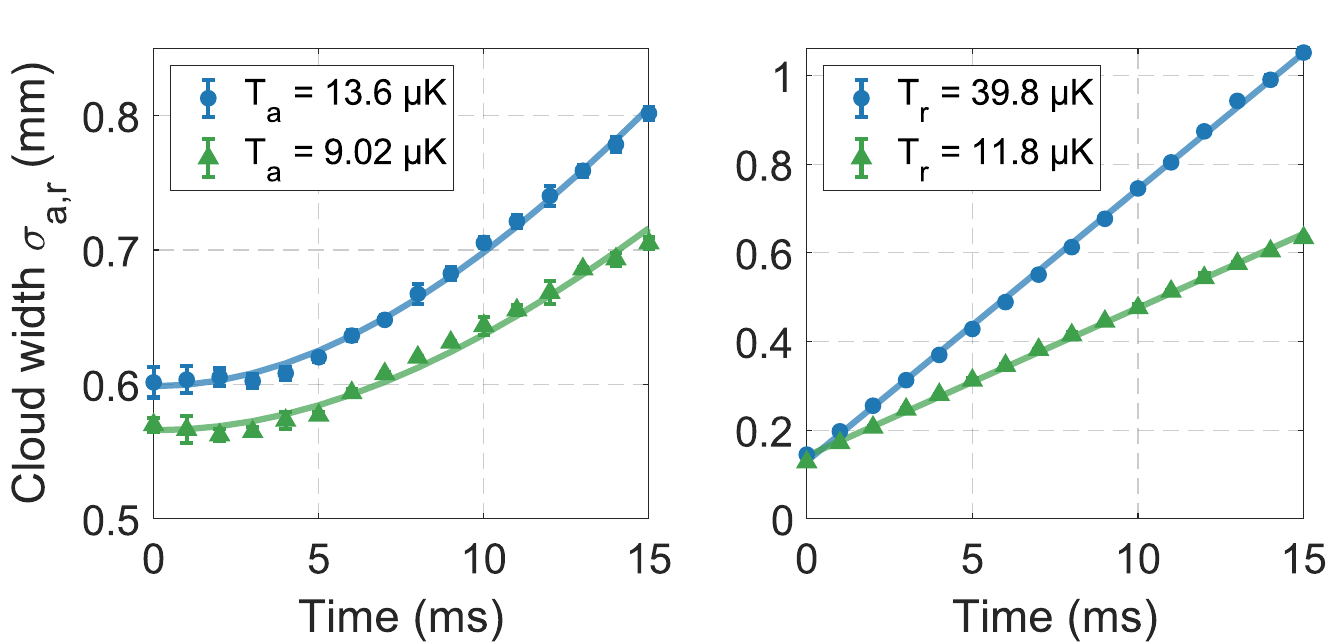}
  \caption{ Measured temperature of the atoms in the axial `a' and radial `r' directions, characterized by a time of flight measurement. Blue circles: Measurements after continuous accumulation. The 1560 nm and 1527 nm intracavity powers were 36~W and 5.2~W respectively throughout accumulation and storage. Green triangles: After accumulation, the 1527 nm is set to 2.8~W and optical molasses is applied for 10 ms.}
  \label{fig:cooling}
\end{figure}

A key feature enabling continuous accumulation is the presence of natural sub-Doppler cooling. We characterize this with a standard time-of-flight measurements~\cite{chu1985three} (Fig.~\ref{fig:cooling}, upper curves). In the axial trapping direction, we measure temperatures as low as 13 $\mu$K, while in the radial direction we obtain temperatures of 40 $\mu$K. Although we do not have a model for this temperature asymmetry, the culprit is the difference in trap frequencies between the two directions; 185 Hz and 0.41 Hz, respectively.
For the extraction of temperatures, we fit the time dependent standard deviation $\sigma(t) = (\sigma_{0}^2 + k_\text{B}T t^2/m)^{1/2}$ for the expanding clouds, with temperature $T$ and initial width $\sigma_0$ as fit parameters; $m$: $^{87}\text{Rb}$ atomic mass, $k_\text{B}$: Boltzmann constant. We note that the cloud width in the radial direction at zero-time should actually be smaller, but we suspect that the diffusion during the resonant imaging process obscures this. Nevertheless, we still observe clear ballistic expansions along the radial direction to extract $T$.

The processes in effect for accumulation and cooling can be understood intuitively. Although the MOT beams are red-detuned by $2.5~\Gamma$ outside of the trap, with the compensation power set to 5.2~W, atoms in the trap are light shifted by $+69 \text{ MHz}$ on average (see Fig. 2), experiencing a red detuning of $14 \:\Gamma$. Thus, as the atoms captured by the MOT near the edge of the trapping potential move toward the center of the dipole-trap, they experience increasingly red-detuned cooling light, starting to undergo polarization gradient cooling~\cite{adams1997laser}. However, instead of this being the standard time sequenced operation, here it is position dependent. Simultaneous MOT capturing and spatially dependent far-detuned cooling reduce the motional energy below the dipole-trap depth, allowing them to be captured in a rather shallow dipole potential. Additional measurements reveal that in steady-state, 85\% of the atoms reside in the $F=1$ manifold, providing further insight: Once the atoms are captured by the dipole trap, they undergo limited cycles of cooling within the $F=2$ manifold. As they move towards the trap center, the repumper becomes off-resonant, and atoms that fall into the $F=1$ manifold are stored in the trap, remaining isolated from the cooling light.

\begin{figure}[t]
  \centering
\includegraphics[width=0.9\columnwidth]{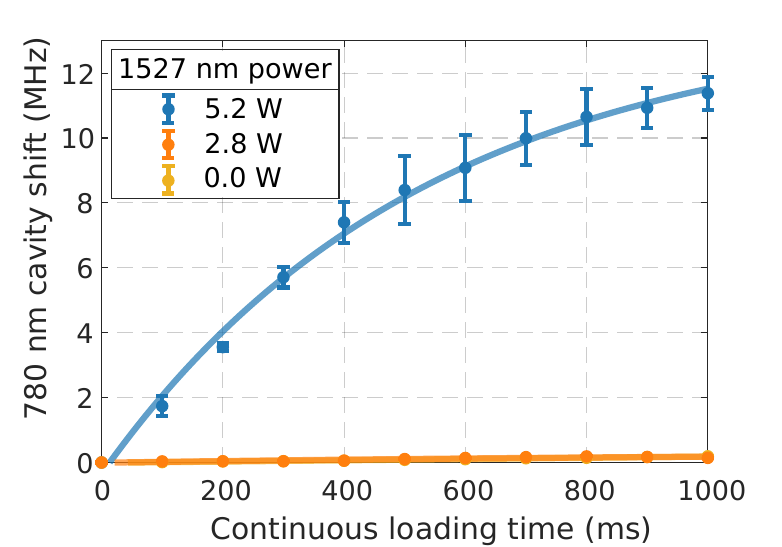}
  \caption{Measured cavity resonance shift for the high finesse 780 nm mode throughout the continuous accumulation process for different light shift compensation levels. Yellow and orange curves are practically indistinguishable. For reference, the full width of the cavity resonance is 85 kHz.}
  \label{fig:cavityshift}
\end{figure}

\emph{Extensions}---We draw attention to the remarkably low levels of the required dipole trap depths to capture the atoms. For example, in Ref.~\cite{engelsen2016quantum}, at least 10 times deeper traps were required to retain a similar number of atoms. In App.~\ref{app: alternative}, we discuss accumulation in even shallower traps, reaching steady-state temperatures of $3~\mu\text{K}$. We further discuss a modified protocol that employs a non-static compensation power capable of additional enhancements in the loading efficiency.

As already noted, after the accumulation of the atoms in the dipole trap, we observe spatially non-isotropic temperatures. It proves straightforward to eliminate this asymmetry after the accumulation phase by further polarization gradient cooling under optimal compensation of the light shifts. This is demonstrated by the bottom curves in Fig.~\ref{fig:cooling}. Here, after accumulation, we set the compensation power to 2.8~W, allowing us to interact with the atoms almost as if the trap did not exist, and run the cooling for 10 ms in zero magnetic field ($- 20 \: \Gamma$ detuning, $0.8 \: \text{mW}/\text{cm}^{2}$ per beam cooling, and $180\: \mu\text{W}/\text{cm}^{2}$ resonant repump). This suggests that there could be optimizations to methods utilized in this work to obtain isotropic velocity distributions also during the continuous accumulation process.

\emph{Continuous strong cavity coupling}---The accumulated atomic ensemble is continuously coupled to the cavity in the strong collective coupling regime during the described protocols, reaching collective cooperativities of order $10^5$ (see App.~\ref{app: parameters}). Such a system can, for example, show continuous vacuum Rabi splitting~\cite{cline2025continuous}. Here we use this continuous coupling in the dispersive regime to observe \emph{in-situ} the atom accumulation in the cavity mode.

Taking into account that the majority of the atoms in the cavity mode are in a specific hyperfine manifold $F$, the total cavity resonance shifts roughly measure the number of atoms in the cavity mode, with around 2~Hz/atom cavity shift (see App.~\ref{app: parameters}). Given their spatially selective and non-destructive nature, these measurements can be inserted at any instant during atom accumulation by scanning the cavity probe beam across the expected resonance. This eliminates the need to hold the atoms in the dipole trap to wait for the surrounding atoms in the MOT to fall before measurements, as was needed for the case of fluorescence imaging. Figure~\ref{fig:cavityshift} shows the cavity shifts observed during accumulation of the atoms for the three specific compensation tone powers, starting with no atoms in the system. The accumulation of atoms towards a steady-state value is clearly seen in the case with the high compensation level. In contrast, for the other two cases, where no atoms remain in the cavity mode at the end, the cavity shifts are much smaller, attaining a relative magnitude of no more than 0.5\%. These residual cavity shifts can then be attributed to the MOT atoms transiting the cavity. 

\emph{Discussion}---Beyond enabling continuous atom–light interfaces, the demonstrated light-shift compensation and continuous loading offer versatile operation for atom–cavity experiments. It becomes possible to perform quantum protocols with static intracavity trapping powers, eliminating the need to ramp trap powers between experimental cycles. This capability is particularly valuable for high-precision sensing, as it allows the cavity mirrors to reach a thermal steady-state and suppresses heating-induced drifts of the cavity resonance. At the same time, the light-shift compensation enables seamless implementation of in-trap molasses cooling, optical pumping for internal-state preparation, and state-dependent removal of atoms from the dipole trap with a resonant push beam. The removal process enables efficient state purification or background measurements with atoms easily removed. All these processes require more considerations and tradeoffs in absence of light shift compensation.

Our results establish a new operational regime in which laser cooling, optical trapping, and cavity coupling coexist seamlessly in steady-state. The demonstrated light-shift-manipulation technique reconciles these normally competing requirements, enabling a stable ensemble of cold atoms to remain continuously coupled to the cavity field. This capability provides the essential experimental ingredient for continuous cavity QED with cold atoms. Because the method relies only on tunable optical parameters, conventional broad-line transitions and very shallow dipole traps, it can be readily adapted to a range of systems. This advance in continuous atom–light interfaces opens new prospects for continuous atomic quantum sensors and hybrid quantum platforms, and offers a path to continuously reloaded atomic quantum processors.

\begin{acknowledgments}
The authors thank Vyacheslav Li for his early contributions to the development of the setup utilized in this work. This work was supported by the Institute of Science and Technology Austria (ISTA).
\end{acknowledgments}

\bibliography{bibliography}
\section*{End Matter}
\appendix
\section{Experimental setup}\label{app: setup}

\subsection{Optical cavity}
As indicated in the main text, the optical cavity was designed for atom interferometry inside the cavity mode with spin-squeezed states and consequently has a number of particular properties: The traveling-wave nature of the cavity is (1) to ensure that the atoms are free to move along the cavity axis in the region of the cavity mode waist and (2) to be able to drive two-photon Raman transitions at 780-nm with counter-propagating beams to impart state-dependent momentum kicks along the cavity mode. The traveling wave cavity is formed by three mirrors in the plane-plane-convex configuration. The convex mirror is mounted on an ultrahigh-vacuum compatible piezo-based translation stage (SmarAct) to allow for in-vacuum fine tuning of the cavity length (Fig.~\ref{fig:setup}). This enables simultaneous accommodation of two 780-nm intracavity tones separated by the 6.834~GHz hyperfine splitting of the $^{87}\text{Rb}$ atoms to drive the Raman transitions.

The cavity mode has a single waist for all wavelengths involved in the setup, with waist values of 157 $\mu\text{m}$, 155 $\mu\text{m}$ and 111 $\mu\text{m}$ for the 1560-nm, 1527-nm and 780-nm modes, respectively, averaged over the two principal (sagittal and tangential) axes. The mode waists are co-located with the crossing region of the MOT beams (Fig.~\ref{fig:setup}). The cavity spacer is built from Macor and the three attached mirrors forming an equilateral triangular mode with a circumference of 9.84~cm.
The custom mirror coatings (Layertec) enhance the difference in the reflectivities of the two polarization modes, resulting in a high-finesse out-of-plane polarization mode ($s$-polarization, FWHM-linewidth $\kappa_{1560,s} = 2\pi \times 30 $ kHz, $\kappa_{780,s} = 2\pi \times 85 $ kHz) and a low finesse in-plane polarization mode ($p$-polarization, $\kappa_{780,p} = 2\pi \times 1.39 $ MHz).
Note that in order to drive cavity-mediated Raman transitions, the cavity has to accommodate fundamental modes with perpendicular polarizations, whose frequency offset matches the hyperfine splitting of \textsuperscript{87}Rb. The polarization dependent phase shifts from the specific mirror coatings also need to be taken into account for identification of the required cavity length to match this offset.

The residual scattering losses in the cavity mirrors can break the strict traveling wave operation by back-scattering light into counterpropagating cavity modes. In the presented experimental setup, we observe a typical back-scattering fraction of $5 \times 10^{-3}$ relative to the power in the primary 1560-nm high finesse mode. The interference between the undesired counter-propagating tone and the primary tone is expected to yield a 14\%-peak residual spatial intensity modulation along the cavity axis for the intracavity trap. Implications for future experiments and potential mitigation techniques are currently under investigation.
  
\subsection{Laser system for cavity interactions}
The 1560-nm source is a few-Hz-level linewidth external-cavity diode laser (OEwaves), the output of which is divided into three branches. The first branch is frequency-doubled in a waveguide doubler (NTT) and coupled to a $\text{TEM}_{00}$ high-finesse 780-nm mode to serve as the cavity probe beam in this work. The second branch is coupled to a $\text{TEM}_{00}$ high-finesse 1560-nm mode after it passes through an electro-optic I/Q modulator in carrier-suppressed single-sideband mode (SSB) \cite{wald2022analog} to form the dipole trap. 
This modulator, operated at 228 MHz, is used for shifting the frequency of the 780-nm tone relative to the 1560-nm mode to compensate for the difference in the effective cavity lengths for the two wavelengths. This configuration ensures simultaneous resonance of the dipole trap tone and the 780-nm tone. 
The third branch, although not utilized in this work, passes through another SSB modulator with a 3.417 GHz shift, after which it is frequency-doubled to serve as one of the future Raman beams, coupled to the low finesse 780-nm $\text{TEM}_{00}$ mode. In the context of Raman transitions, the probe beam serves as the second arm, completing the two-photon transition. Finally, the light shift compensation tone is obtained from a separate 1527 nm fiber laser (NP-Photonics) with a linewidth of 500~Hz. This light is directly coupled to a 1527-nm counter-propagating low-finesse $\text{TEM}_{00}$ mode. 

We apply squash lock \cite{diorico2024laser} for the 1560-nm laser-cavity lock. 
Squash lock is a modulation-free approach which uses a quadrant photo-diode (QPD) to derive an error signal from changes in beam shape ellipticity in cavity reflection. 
The cylindrical lens pair in Fig.~\ref{fig:setup} serves for the beam shaping which is essential for this method. The analog error signal is filtered \cite{fox20031} and distributed to three feedback channels:
High frequency noise is suppressed by feedback to the 1560-nm laser, which is simultaneously keyed to a\textsuperscript{85}Rb transition via a low-bandwidth offset lock. Low frequency drifts ($< 10 \text{ Hz}$), and structural resonances ($\sim$3.5 kHz) are suppressed via feedback to the linear stage and ring piezo, respectively.

\subsection{Laser system for MOT}
The MOT laser system generates the cooling and repump tones addressing the $\ket{5\text{S}_{1/2}, F = 2} \leftrightarrow \ket{5\text{P}_{3/2},F'= 3}$ and $\ket{5\text{S}_{1/2}, F = 1} \leftrightarrow \ket{5\text{P}_{3/2},F'=2}$ transitions of $^{87}\text{Rb}$ \cite{steck2001rubidium}. These tones are derived from 780-nm external-cavity diode lasers (Sacher) and are frequency-offset locked \cite{li2022laser} to a reference laser that is itself locked to the $^{85}\text{Rb}$ transition $\ket{5\text{S}_{1/2},F=3} \leftrightarrow \ket{5\text{P}_{3/2},F'=4}$ via modulation-transfer spectroscopy \cite{raj1980high}. In typical operating conditions, each of the three pairs of counter-propagating MOT cooling beams has an intensity of $10~\text{mW}/\text{cm}^{2}$ with a spot size of 0.8~cm and a red-detuning of $-2.5\:\Gamma$. There is a single beam for the repump tone with an intensity of $5~\text{mW}/\text{cm}^{2}$ with a spot size of 0.8~cm.
The magnetic field gradient for the MOT is $10~\text{G}/\text{cm}$. A 2D-MOT-based atom source (Cold Quanta) delivers Rubidium atoms into the science chamber through a pinhole. 
The loading location of the MOT is aligned with the cavity mode waist and can be adjusted within a couple of millimeters through external magnetic bias fields. 
In typical operating conditions, the MOT loads at a rate of $6\times10^{7} ~\text{atoms}/\text{s}$ and maintains the loaded atoms at temperatures around $300 \: \mu \text{K}$, as indicated by time-of-flight measurements. We measure the $1/e$ lifetime of the MOT to be 5.4 s. Fluorescence imaging (FLIR Blackfly CMOS camera) is used to determine the atom numbers and temperatures. The conversion of detected counts to atom numbers for the fluorescence based atom number calibration can be done as in Ref. \cite{steck2001rubidium}. However, for this atom number calibration, we use a cavity-based calibration method, more suited for our application; see App.~\ref{app: parameters}.

\subsection{Additional peripherals}
 A microwave antenna is integrated inside the vacuum chamber in the form of a $\lambda/4$ dipole antenna that is directly attached to an SMA vacuum feed-through. 
The antenna is used for coherent manipulation of the atomic state between the \textsuperscript{87}Rb magnetically insensitive hyperfine clock states, separated by 6.834 GHz. 
 For driving this antenna, a 7.2-GHz ultra-low noise microwave source (Wenzel Assoc.) is frequency mixed with an agile direct digital synthesizer (AD9910) operated at 366 MHz. The Rabi oscillation rates induced by this source is $\Omega_R/2\pi=3.8~\text{kHz}$ with 28 dBm of input microwave power. This spectroscopic tool is utilized, for example, for zeroing the magnetic fields at specific instances in this work by observing the magnetic Zeeman shifts on the ground state sub-levels.

For the atomic state preparation, we optically pump the atoms into the magnetically insensitive $\ket{F = 1, m_F=0}$ state.
This is achieved by driving the atoms with a $\pi$-polarized repump tone tuned to the $\ket{5\text{S}_{1/2}, F=1} \leftrightarrow  \ket{5\text{P}_{3/2}, F'=1}$ transition, while the cooling light is tuned to the  $\ket{5\text{S}_{1/2}, F=2} \leftrightarrow  \ket{5\text{P}_{3/2}, F'=2}$.
In this configuration, the $\ket{F = 1, m_F=0}$ state is a dark state due to the selection rules, where the atoms accumulate~\cite{avila1987state,duan2017state}.
 
The experimental sequences are controlled through the ARTIQ Hardware, using the Quantrol open-source graphical user interface developed in our group \cite{Quantrol2025}. 

\section{Light shifts} \label{app: shifts}

\begin{figure}[t]
  \centering
\includegraphics[width=0.9\columnwidth]{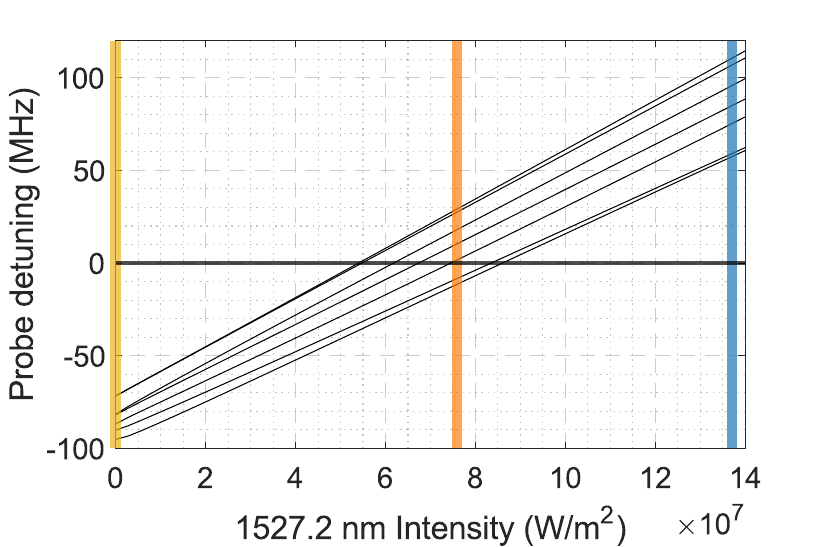 }
  \caption{Numerical calculation of the light shifts for the \mbox{$5\text{P}_{3/2}$, $F'=3$} magnetic sublevels as a function of compensation power, showing a light-induced lifting of the degeneracy of the sublevels in addition to a linear mean shift. The zero-shift level is that of an atom free from both the trap and compensation lights. The three compensation intensities used repeatedly in the main text are shown, and the trap tone intensity matches that in the main text.}
  \label{fig:tensorshift}
\end{figure}

In the main text, level shifts due to interactions with linearly-polarized far-off-resonant light has been treated utilizing \emph{scalar} polarizabilities. In this approximation, all magnetic sublevels associated with a hyperfine level experience identical shifts. However, it is well know that \emph{tensor} polarizabilities lift this degeneracy \cite{arora2007magic,safronova2011critically} resulting in a spread of sublevel shifts ~\cite{shih2013nondestructive}, which can even  become non-perturbatively large for light with small detunings \cite{Coop17}. Depending on the relative polarizations of the trap and compensation lights, a compensation of the scalar shifts need not indicate a compensation of the tensor part, raising the question as to what extent the tensor shifts might interfere with laser cooling. A numerical analysis (using the code provided in Ref.~\cite{Coop17}) for orthogonal linearly polarized trap and compensation lights shows that, in this configuration, the tensor shifts indeed do not cancel (Fig.~\ref{fig:tensorshift}) and always lift the degeneracy of the \mbox{$5\text{P}_{3/2}$, $F'=3$} magnetic sublevels: A full sublevel spread of 25~MHz, 40 MHz and 55~MHz for the zero, intermediate and high compensation powers utilized in the main text. Note that this analysis does not take into account the spread in the light shifts due to the spatial distribution of the atoms in the dipole trap at finite temperatures.

Spectral shapes associated with light shift distributions in dipole traps have been studied in various configurations~\cite{Coop17,shih2013nondestructive}. For our purposes, simple Lorentzian fits prove sufficient to extract approximate full-widths yielding 40~MHz, 25~MHz and 80~MHz for the zero, intermediate and high compensation levels for the distributions in Fig.~\ref{fig:energy_shifts}. Note that these distributions are measured for atoms at 11~$\mu\text{K}$ in both axial and radial directions following an additional molasses stage after accumulation, and that the power-broadened linewidths in absence of any trap or compensation tone read 10 MHz for the utilized probe power. We find that the widths corresponding to the zero and high compensation powers show a more pronounced dependence on the radial temperatures (not shown) in comparison to the width corresponding to the intermediate compensation power where the mean shifts are canceled. This indicates that the spatial dispersion in the shifts are also a significant part of the observed widths unless the mean shift is canceled.

The combined effects of the tensor shifts and spatially broadened light shifts provide a picture that correctly reproduces the hierarchy of the linewidths observed in Fig.~\ref{fig:energy_shifts}, albeit without an exact quantitative agreement. Nevertheless, further developing a model that can better predict the observed linewidths goes orthogonal to the scope of the current work, as the magnitude of the spread in the energy levels either due to the simulated tensor shifts or spread in the radial positions seem not to be relevant for capturing and cooling the atoms in the intracavity dipole trap given the relatively large effective detunings for the cooling light in the intracavity region---see main text.

\section{Spin dependent cavity shifts}\label{app: parameters}
We calibrate our fluorescence-based atom counting using the expected cavity shifts per spin-flip given the well characterized nature of the quantities that go into evaluating the atom-cavity coupling. 

The quantization axis for the atoms in the dipole trap, defined by a 70-mG external magnetic field, is aligned parallel to the polarization of the 780-nm high-finesse cavity mode, coupling this mode to the atoms with \mbox{$\pi$-polarization}. Focusing our attention to the two magnetically insensitive states $\ket{F = 1, m_F=0}$ and $\ket{F = 2, m_F=0}$, the cavity length is set such that the cavity mode couples these two states to the $5\text{P}_{3/2}$ excited state manifold with opposite detunings of $\Delta=\pm3.417$, corresponding to half the hyperfine splitting frequency. The effective interaction part of the Hamiltonian for this collective two-level system in this dispersive configuration is 
\begin{equation}
    \hat{H}_{\text{int}} = \hbar \Omega \hat{a}^\dagger \hat{a} \hat{J}_z
    \label{eq:effhammy}
\end{equation}
where $\Omega = 2g^2/\Delta$ is the dispersive frequency shift, $2g$ is the vacuum Rabi splitting and $\hat{J}_z$ is the collective spin population difference operator \cite{schleier2010squeezing,hammerer2004light}. Given the cavity mode parameters, the polarization and the associated atomic transition matrix elements, the theoretical atom-cavity coupling rate and the cavity shift per spin flip are evaluated to be $g = 2\pi\times85~\text{kHz}$ and $\Omega/2\pi=4.2~\text{Hz}$ respectively. The resulting single-atom cooperativity parameter is $C = 
{4g^2}/{\kappa \Gamma}= 0.056$ for this high finesse 780-nm traveling-wave mode. With $N=4.0\times10^6$ atoms loaded into the dipole trap (see App.~\ref{app: alternative}), collective cooperativity can reach $NC=2.2\times10^5$, putting the system deeply into the strong collective coupling regime.

For the atom number calibration, after loading atoms into the dipole trap and further cooling, we state prepare them in the $\ket{F = 1, m_F=0}$ ground state through optical pumping and a consecutive state-selective removal of the atoms from the trap with the push beam to purify the spin state. Following this state preparation, we observe 98.7$\%$ contrast microwave-driven Rabi oscillations via fluorescence imaging. We calibrate the observed fluorescence signals using the independently measured cavity resonance shifts during the Rabi oscillations.

For the cavity shifts measured during continuous atom accumulation (Fig. \ref{fig:cavityshift}), since the majority of atoms are observed to be in the $F=1$ manifold, a 2~Hz/atom conversion factor can be taken as a guide to estimate the number of atoms located in the cavity mode. However, note that this is just a guide, since atoms are distributed across different magnetic  manifolds leading to different cavity couplings, and the exact spatial distributions of the atoms are not known during the loading procedure.

\section{Alternative loading schemes}\label{app: alternative}

Here we briefly mention a couple other loading variations we tried that can optimize different quantities of interest. 

First, we remark that even shallower traps can be utilized to continuously accumulate atoms, significantly relaxing the required laser powers compared to standard methods. This, in fact, optimizes for low temperatures with marginally reduced atom numbers. We have observed continuous accumulation in a 35-$\mu\text{K}$-deep trap (10 W), with atom numbers exceeding $1.5 \times 10^6$ at $3~\mu\text{K}$ longitudinal and $9~\mu\text{K}$ radial temperatures in steady state.

For loading even more atoms into our usual 87-$\mu\text{K}$-deep trap, we have devised a loading scheme, albeit one that deviates from the purely `steady-state' operation principle. In this scheme, for the first 500 ms, we set the 1527 nm intracavity power to 2.8~W and MOT parameters to full power (see App. \ref{app: setup}). This allows the MOT to collect atoms as if in free space. Then, we apply a linear ramp of the 1527 nm power to 5.2~W in the last 50 ms. This results in $4 \times 10^6$ atoms ($12\%$ of the MOT) loaded into the dipole trap with axial and radial temperatures similar to those discussed in the main text. This constitutes the highest reported percentage of MOT to 1560-nm-dipole trap transfer efficiency.
However, it shall be noted that higher efficiencies have previously been reported for other trap laser wavelengths~\cite{kuppens2000loading,brzozowski2010diagnostics}. Although this alternative method breaks the `steady-state' operation, the change in intracavity power is minimal: $\Delta P_{1527} = 2.5 \: \text{W}$ in comparison to the 36~W trap power, leading to negligible photothermal effects.

\end{document}